\documentclass[biblatex]{lni}
\usepackage{booktabs}

\usepackage[]{blindtext}

\begin{document}
\title[Ein Kurztitel]{Object classification with Convolutional Neural Networks: from KiDS to Euclid}
\author[1]{G.A.~Verdoes Kleijn}{}{}
\author[2]{C.A.~Marocico}{}{}
\author[3]{Y.~Mzayek}{}{}
\author[4]{M.~Pöntinen}{}{}
\author[5]{M.~Granvik}{}{}
\author[6]{O.~Williams}{}{}
\author[7]{J.T.A.~de Jong}{}{}
\author[8]{T.~Saifollahi}{}{}
\author[9]{L.~Wang}{}{}
\author[10]{B.~Margalef-Bentabol}{}{}
\author[11]{A.~La Marca}{}{}
\author[12]{B. Chowdhary Nagam}{}{}
\author[13]{L.V.E.~Koopmans}{}{}
\author[14]{E.A.~Valentijn}{}{}
\affil[1]{University of Groningen, Groningen, The Netherlands}
\affil[2]{University of Groningen, Groningen, The Netherlands}
\affil[3]{University of Groningen, Groningen, The Netherlands}
\affil[4]{University of Helsinki, Finland}
\affil[5]{University of Helsinki, Finland and Lule\aa{} University of Technology, Lule\aa{}, Sweden}
\affil[6]{University of Groningen, Groningen, The Netherlands}
\affil[7]{University of Groningen, Groningen, The Netherlands and Leiden Observatory, Leiden, The Netherlands}
\affil[8]{University of Groningen, Groningen, The Netherlands}
\affil[9]{SRON Netherlands Institute for Space Research, Groningen, The Netherlands and University of Groningen, Groningen, The Netherlands}
\affil[10]{University of Groningen, Groningen, The Netherlands}
\affil[11]{University of Groningen, Groningen, The Netherlands and SRON Netherlands Institute for Space Research, Groningen, The Netherlands}
\affil[12]{University of Groningen, Groningen, The Netherlands}
\affil[13]{University of Groningen, Groningen, The Netherlands}
\affil[14]{University of Groningen, Groningen, The Netherlands}
\maketitle

\begin{abstract}
Large-scale imaging surveys have grown $\sim$1000 times faster than the number of astronomers in the last 3 decades. 
Using Artificial Intelligence instead of astronomer's brains for interpretative tasks allows astronomers to keep up with the data. 
We give a progress report on using Convolutional Neural Networks (CNNs) to classify three classes of rare objects (galaxy mergers, strong gravitational lenses and asteroids) in the Kilo-Degree Survey (KiDS) and the Euclid Survey.
\end{abstract}


\section{Needles in the Kilo-Degree Survey and the Euclid Survey haystacks}

\begin{figure}
    \centering \includegraphics[width=\linewidth]{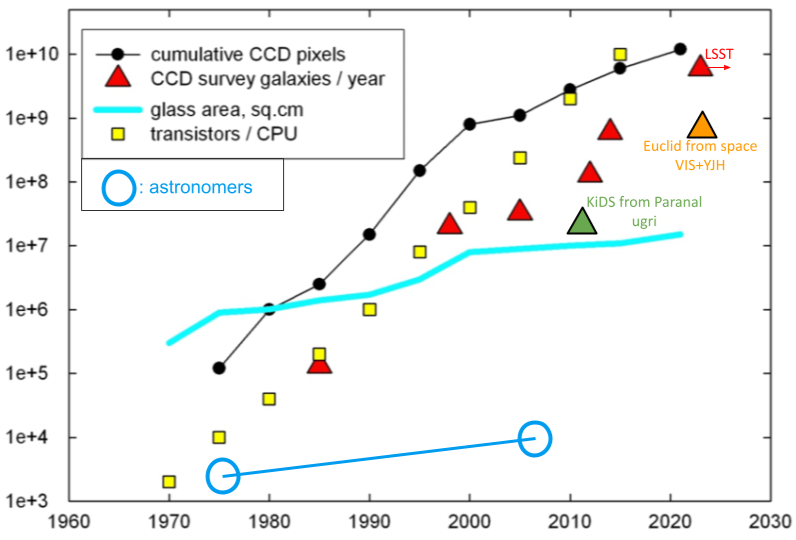}
    \caption{Growth of information technology, galaxy surveys and astronomers as a function of time.} 
    The typical number of transistors per CPU, the aperture area of telescopes ("glass"), the cumulative number of pixels in telescope cameras and number of astronomers are shown. The plot is adapted from \citep{tyson19}. The triangles show for surveys the total number of galaxies in any filter with S/N $>$ 5 for flux inside a 2 arcsec wide aperture per year. The arrow for the LSST triangle indicates it is currently planned to start late 2024.
    \label{HaystackGrowth}
\end{figure} 

\begin{figure}
    \centering
    \includegraphics[width=\linewidth]{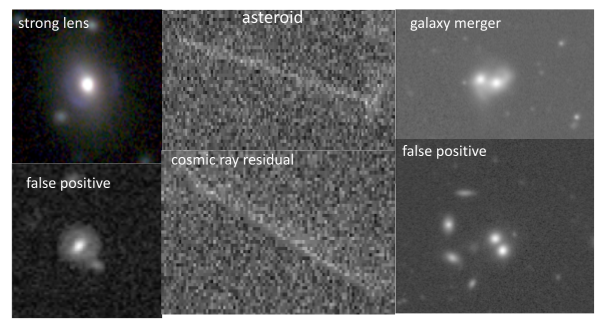}
    \caption{True positives for the three classes of objects in top row and false positives in bottom row. The thumbnails are from KiDS for the classes of strong lenses and galaxy mergers and from simulated Euclid data for the asteroid class.}
    \label{TrueAndFake}
\end{figure} 


Figure~\ref{HaystackGrowth} shows that observational digital information on galaxies has grown roughly 1000 times faster than the number of astronomers since 1985.

The figure is taken from \citep{tyson19} and data points for the Kilo-Degree Survey \citep{kuijken19} and the Euclid Survey \citep{scaramella22} have been added. 
The development of information systems and their algorithmic software to analyze these digital data has been critical to accommodate this growth. 
Information systems have taken over many operational tasks performed earlier by astronomers. The next step is that systems will take over from astronomers interpretative tasks such as the classification of celestial objects based on their morphology, color and context in images.

In the recent past classification relied on human judgement via the careful, laborious creation of both customised catalogs with up to hundreds of features extracted from pixels and also customised algorithms that took these features as input and gave a classification as output. 
CNNs offer the option to let information systems go from pixels straight to classification in a scalable, reproducible and objective manner. 
They bypass the route of, possibly subjective, human interpretation and the laborious task of feature cataloging and classification algorithm creation.

Our teams of astronomers and data scientists focus on the detection and classification of three classes of rare objects in KiDS and Euclid: gravitational strong lenses, galaxy mergers and asteroids. 

We expect 2.4E3 / 1E5 strong lenses, 1E4 1E4 / 1E7 galaxy mergers and 6.4E4 / 1.5E5 asteroids in KiDS / Euclid respectively.

They are the "needles in the haystack" of images. 
Only of the order of 1 in 1000 massive galaxies are a strong lens, only of the order of 10\% of galaxies are in major mergers, while the number of asteroids per square degree and hence of the order of 1E5 source detections can be as low as a few. 

For each class there are many more sources that resemble these classes, but are "impostors" (e.g., Fig~\ref{TrueAndFake}).

\section{Classification with convolutional neural networks}

CNNs have emerged as a powerful approach to classify massive datasets of source images because they can learn from their own feature extraction and optimise it for the desired classification (e.g., \cite{huertas22} for a review). The feature extraction that is acquired through training for the properties of a particular imaging survey can be transferred to other imaging surveys using transfer learning. This makes the approach potentially very generalisable.   

The first CNNs deployed to find strong lenses in KiDS had a ResNet architecture (\cite{petrillo17}, \cite{petrillo19a}, \cite{petrillo19b}, \cite{li20}, \cite{li21}). We have currently developed a “DenseLens” approach based on a DenseNet architecture \citep{nagam23}. The DenseNet architecture uses concatenation of all preceding feature maps and this improves the flow of information from the first layer until the last layer. This also helps to reduce the total number of parameters in the network by an order of magnitude.

Each training step has an equal number of lenses and non-lenses. Simulated lensing features were generated and added to thumbnails of elliptical galaxies from KiDS to create lenses. Non-lenses are a combination of hand-picked "impostors" and elliptical galaxies. In the training the Binary Cross Entropy loss were minimized using an ADAM optimizer. The approach includes a technique to rank-order strong lenses without human inspection by training on a metric for the information content of the lensing features in the training data. The automated rank-ordering is important for the Euclid Survey given the expected 10$^5$ true strong lenses from it. 

The CNNs deployed on KiDS for galaxy mergers used transfer learning from a pre-trained VGG19 CNN \citep{wang20}. Given that differences in appearance between merging galaxies and normal galaxies  can be subtle it is crucial to have a comprehensive training sample. This is why hydrodynamical simulations are used to construct our training sample. The 3D merging galaxies in the simulations are converted to mock observations for the particular survey. Currently a new custom CNN is being trained for survey data from the Hyper Suprime Camera on Subaru as preparation for Euclid. Early results show an accuracy of 77 per cent, with reliability on merger detection of 83 per cent.

The CNN for asteroids has six convolution layers and a desne layer. The training is done with a YOLO loss function. For KiDS, the output layer has one unit outputting a classification label with a sigmoid function. For KiDS the training data consists of true asteroids in KiDS detected using a classical detection pipeline \citep{mahlke18} and other detections as non-asteroids. For Euclid the training examples are based on simulated Euclid VIS images generated with the Euclid Consortium's \texttt{ELViS} software with added simulated asteroid streaks (courtesy L.~Conversi and A.~Nucita). For Euclid, the training includes an Adam optimizer \citep{kingma2014} and early stopping callback monitoring the validation loss. For Euclid, the machine learning pipeline \citep{pontinen23} surpasses the asteroid detection completeness (recall) achieved by a traditional segmentation-based pipeline \citep{pontinen2020}, detecting both fainter and slower-moving asteroids.

\section{Next steps}

It is clear from the results published and from the work in progress discussed above for KiDS and Euclid that CNNs provide a powerful way to address the interpretative task of classification in current and future astronomical surveys . 

The workflow of deploying the CNNs above on surveys still includes the creation of feature catalogs for detecting sources and classifying stars and extended objects. To bypass the creation of any catalog for all three classes we are experimenting to prepend the workflow with a discriminator CNN that makes preliminary classification of objects into these three classes. The output of the discriminator is then fed to the class-specific CNNs. Initial results maintain 100\% completeness and more than 50\% purity in the output ensembles using class weighting as hyperparameter.   

Next steps also include attempting to let CNNs perform other types of interpretative tasks.     
For example, for the asteroid CNN for the Euclid Survey there is a more advanced version of the model under development, which has five output units and a tailor-made loss function directly predicting asteroid streak coordinates, and which can be used to detect and locate asteroids in full CCDs with a sliding-window algorithm \citep{pontinen23}. This approach thus bypasses not only the need for any detection catalog it also performs an additional interpretative in addition to classification: astrometry.

\printbibliography

\end{document}